# Multistate nonvolatile straintronics controlled by a lateral electric field

V Iurchuk, B Doudin and B Kundys

Départment Magnétisme des Objets NanoStructurés, Institut de Physique et Chimie des Matériaux de Strasbourg (IPCMS), UMR 7504 CNRS-UdS, 67034 Strasbourg, France

E-mail: kundys(A)ipcms.unistra.fr

**Abstract**
We present a multifunctional and multistate permanent memory device based on lateral electric field control of a strained surface. Sub-coercive electrical writing of a remnant strain of a PZT substrate imprints stable and rewritable resistance changes on a CoFe overlayer. A proof-of-principle device, with the simplest resistance strain gage design, is shown as a memory cell exhibiting 17-memory states of high reproducibility and reliability for nonvolatile operations. Magnetoresistance of the film also depends on the cell state, and indicates a rewritable change of magnetic properties persisting in the remnant strain of the substrate. This makes it possible to combine strain, magnetic and resistive functionalities in a single memory element, and suggests that sub-coercive stress studies are of interest for straintronics applications.

Keywords: piezoelectrics, ferroelastics, magnetostriction

## 1. Introduction

The ever increasing miniaturization and low power consumption needs of electronics motivate the search for new possibilities for controlling the magnetic state of spintronic devices. More generally, enhanced memory devices would greatly benefit from multi-level storage and multi-stimuli excitations for writing a memory cell. Electric field is among the preferred approaches for external memory control [1]. Low-power electrical control of magnetization offers the opportunity of combining the individual advantages of MRAMs and FeRAMs in novel types of electrically controlled nonvolatile magnetic bits. A common property fulcrum can be a cross-functional magnetoelectric material [2], provided large enough and non-volatile coupling exists. An alternative approach is to use strain as a common physical property linking magnetic and electrical orders [3] in artificial magnetoelectric structures. Strain is also becoming an increasingly popular method for electrical control of magnetization [4–11], magnetoresistance [12] or magnetic domain wall propagation [13–15]. The recent claims of low power consumption [16, 17] in such systems make strain-controlled devices possible contenders for future logic circuit design. However, most successful reports on piezoelectric magnetization control rely on measurements under large applied electric fields significantly exceeding the ferroelectric coercive values. The large strain created by poling a ferroelectric material beyond its coercive field is considered technically unattractive [18], leading to fatigue and instability of the device. We are aware of a single successful report in the literature where a non-volatile change of magnetization was found using sub-coercive applied electric fields [19], but this required the use of a large piezoelastic stack with an electric field applied to the whole structure, ensuring a significant bulk elongation for realizing a two-states strain. Here we propose a multistate operation in a lateral electrodes design compatible with planar 2D structures. This is complementary to the usual vertical stacks used for ferroelectric memory cells and provides extra room for designing 3D miniaturized architectures. We show that the resulting strain configuration can effectively result in multistate magnetic anisotropy control and multistate electroresistance control at low sub-coercive applied electric fields, making the device attractive for low-power and reliability performances. Moreover, we are able to re-instate the initial magnetic and resistive properties of the magnetic overlayer by recovering the initial strain equilibrium of the substrate surface. We attribute these remanence properties to the ferroelastic-like behaviour of the substrate, with a remnant strain persisting even though the applied electric field is below the coercive poling value [20]. PZT is among the ferroelectric materials that change crystal class at their Curie point, and can exhibit ferroelastic properties [21, 22]. The related 90° domains (commonly called ferroelastic domains in PZT) are usually considered as unfavourable for piezoelectric and ferroelectric applications. We are aware of one report of flexoelectric properties [23] where the stress gradient provides control of the ferroelectric domains [24]. There is however a growing interest in better understanding of the domains and domain walls in ferroelectric materials owing to their importance for devices and miniaturization issues [25, 26]. Here, we show how tunable strain hysteresis can result in multistate resistance changes of the film deposited onto piezoelectric ceramics, demonstrating a pure electoresistive rewritable memory effect applicable to ferromagnetic materials, and possibly extended to spintronics functionality.

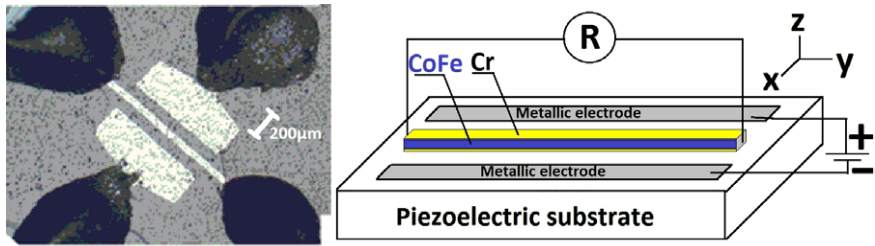

**Figure 1.** Microscopic picture (left) of the sample and schematics of experiment (right).

## 2. Results

Figure 1 shows the microscopic picture and the schematics of the planar ferroelastic cell. In a view of the possible integration with hybrid straintronic-spintronics, we tested $Co_{50}Fe_{50}$ magnetoelastic film as top active layer material. We chose a commercially available ceramic of lead zirconate titanate (PZT) [27] as a substrate, which is among the most common and widely used piezoelectric materials [28]. The PZT (ref: PIC 255) ceramics was received with Ag electrodes that were removed by polishing without postannealing. A 70 nm thick polycrystalline CoFe film with 10 nm Cr adhesion layer was deposited in an e-beam evaporator, with base pressure of $4e^{-8}$ mbar and at a rate of 0.2 nm s$^{-1}$. A top 3 nm of Cr was deposited to prevent oxidation of the magnetic layer. We used a shadow mask for patterning a CoFe line (34 $\mu$m wide and 960 $\mu$m long along the *y axis* of figure 1), providing an electroresistive response related to the strain-induced geometrical changes of the stripe [29]. The electrodes were connected using silver paste and the resistance of the film was measured by a LCR meter. For the purpose of presenting a proof-of-principle experiment with a straightforward measurable signal, we restricted ourselves to the simplest strain gage design. As shown in figure 2, the resistance of the deposited film shows a remarkable hysteresis. The electric field is applied to an initial state sample with resistance R1, with remnant memory states of resistance values R2 to R13 measured at subsequent zero applied field.

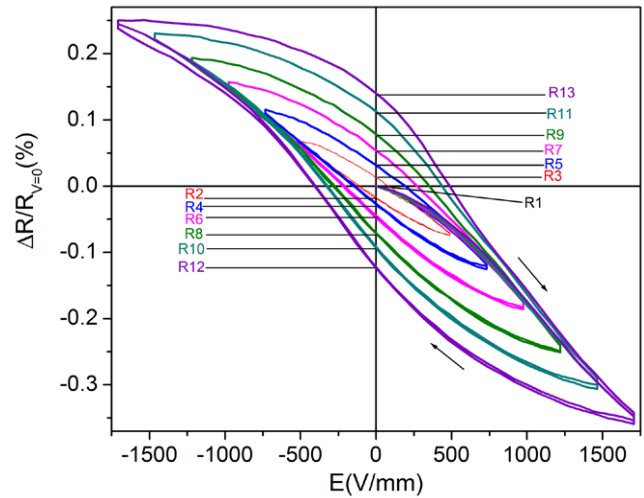

**Figure 2.** Electroresistive hysteresis of CoFe film with multi memory states.

The switching is very reproducible upon cycling, with testing of $10^7$ cycles without detecting fatigue. We attribute this stability to the applied electric fields significantly below the ferroelectric coercive force, allowing us to build many stable memory states by voltage pulse amplitude and sign simple manipulations, without poling the substrate layer. This data therefore clearly indicates the possibility to multi-state, or analog storage of the resistance state, in contrast to the binary information storage typical of a standard memory cell. Figure 3 confirms the stability and diversity of the multiple resistance states created by this technique. Writing voltage pulses of approximately 100 ms duration of different amplitude (figure 3(*b*)) were used. A small relaxation after the pulse can be observed, with distinct stabilized values categorized as r1-r17 after a few seconds. Note the slow dynamics of the resistive reading, possibly indicating mechanical instabilities in the system. Such effect, and in particular how it evolves when miniaturizing the cell, remain to be investigated and is beyond the scope of this letter. The number of presented resistance states values is limited here by the resolution in the resistance measurements and the slow dynamics of the post-pulse resistance drift. We recall that the data of figure 2 is obtained without optimization of the strain gauge material, geometry, dimensions, and PZT surface quality.

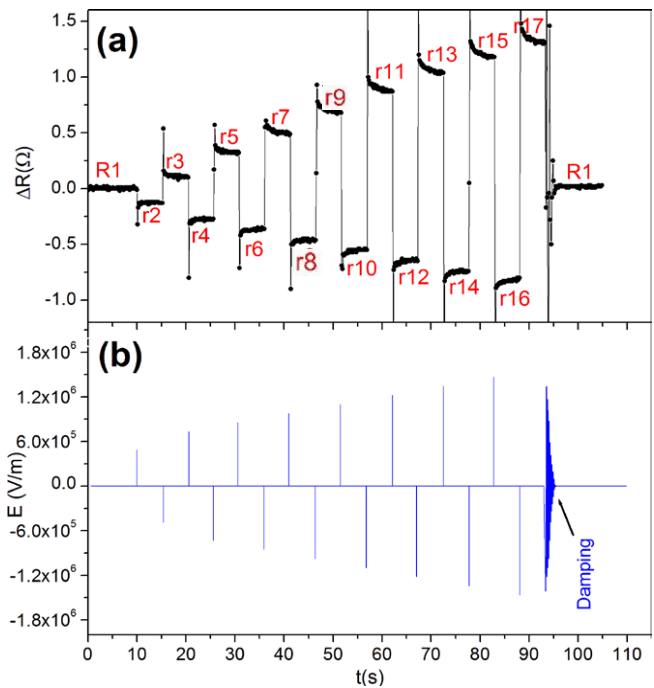

**Figure 3.** Electric field induced switching between multiple memory states (*a*) with a corresponding electric field time profile (*b*). Note that the lettering of these states is different from the one in figure 2 to distinguish the different history of electrical writing of the states.

Although it is rather difficult to recover the exact domain configuration of the virgin state, the corresponding deformation equilibrium and the corresponding initial resistance of the film can successfully be recovered on-demand by applying a damping voltage procedure with a decreasing and oscillating time profile shown in figure 3(*b*). Therefore, an additional 'erased' memory state R1 can be formed and recovered on-demand. We also observe that the imprinted resistance states can remain stable over weeks, confirming the interest of the low stress fields approach for memory applications. The multiplicity of the possible remnant strains likely relates to the multiplicity of 90° ferroelastic domains in PZT, with displacements and pinning/unpinning occurring at small electric field stress values. We note here that the domain walls dynamics in ferroelectrics are rather well documented for large applied fields, but the sub-coercive behaviour is essentially unchartered [18]. The only exception we are aware of relates to the observed bowing of the pinned domain walls [30] that can contribute to the electromechanical properties in [31, 32]. Although, the question of scaling down the size of the device cell remains open, one can nevertheless expect that sub-100 nm domains structures can be stabilized in thin films and controlled in micron-sized cells [25, 26, 33], making a multi-state approach a possible way to circumvent the difficulties in stabilizing the polarization states of FE cells when miniaturizing the device [34]. Moreover by reducing the size of the device, changes in resistance of several k$\Omega$ can easily be expected.

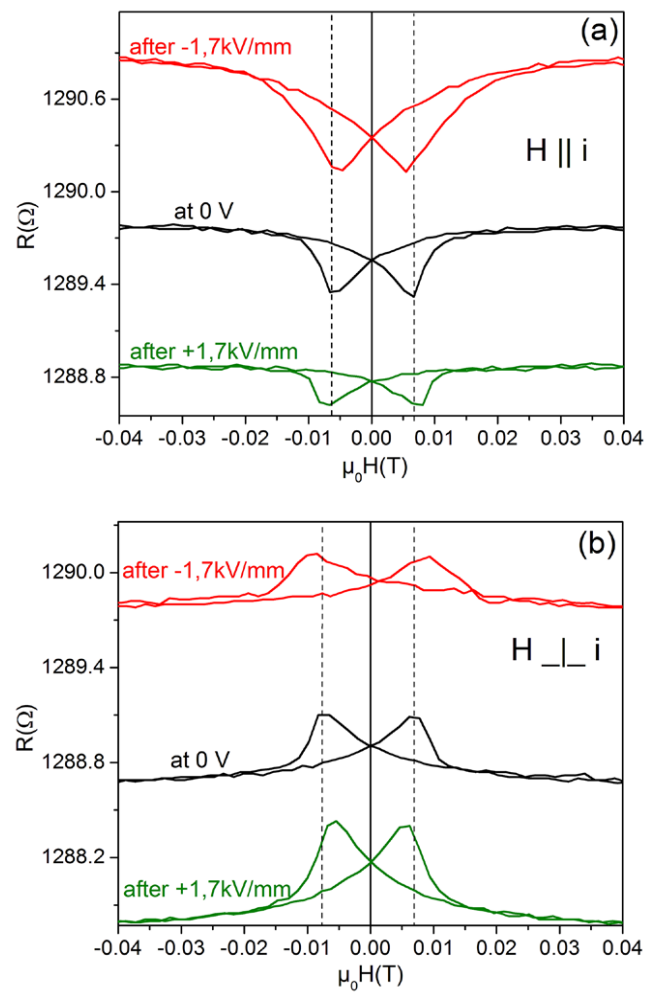

**Figure 4.** Magnetoresistance curves of Cr(10 nm)/CoFe(70 nm)/Cr(3 nm) stripe after removal of the indicated electric field stress values, with applied magnetic field (*a*) parallel and (*b*) perpendicular to the electrical current in the stripe.

To further test if magnetic properties are affected by the strain remnant states, we measured magnetoresistive (MR) loops starting from the 'virgin' state (R1 in figure 2) and strain remnant states observed after voltage pulses corresponding to $\pm 1.7$ kV mm$^{-1}$ electric fields (figure 4). Note that the resistance states indicated in figure 4 no longer correspond to the remnant states of figures 2 and 3, as the observed resistance value is a combination of the initial electroresistance modified by an external magnetic field changing the magnetic configuration of the stripe, and therefore its resistance value.

When the applied magnetic field is parallel or perpendicular to the current direction (figure 4 ), a magnetoresistance curve indicative of the anisotropic magnetoresistance response of the stripe is observed, with corresponding minima in the loop expressing the maximum deviation of the magnetization from the saturated configuration [35]. Notably the loop reaches its saturation at 0.015 T, as expected for $Co_{50}Fe_{50}$ films [36]. After a positive voltage pulse of ~1.7 kV $mm^{-1}$, the magnetoresistance decreases in amplitude and becomes wider (the minima resistance now happen at higher magnetic field values). On the contrary, a larger magnetoresistance is found after a negative voltage pulse, with the noticeable loop shrinking (the points corresponding to minimum resistance now happen at lower magnetic fields as compared to the 0V middle loop). Such behaviour illustrates a strain induced change of magnetic properties, indicating a change in magnetic anisotropy under electric field. This is confirmed by data when the applied field is transverse to the current (figure 4(b)).

While the resistance scale is globally shifted between figures 4(a), (b) due to the anisotropic magnetoresistance of CoFe, the shapes and amplitudes of the magnetoresistance curves clearly depend on the applied electric field pulses. These changes in figure 4(b) are reversed when compared to figure 4(a). This indicates, in first approximation, that the easy magnetization direction is strain-dependent [19]. Importantly, we are able to fully recover the initial magnetic states for both perpendicular and parallel configurations by using the simple damping electric field procedure (figure 3(b)).

## 3. Conclusions

On-demand change of the remnant strain configuration of a ferroelectric substrate, with deterministic control and a multiplicity of the possible states, has therefore been successfully obtained. This provides a 'room inside' for electroresistive memory devices operating at low power, with remarkable multistate non-volatile operation. Moreover, the possible integration of the effect reported here with spintronics elements leads to an additional degree of freedom in hybrid straintronic-spintronic devices where multiple and stable changes in magnetization can be created at low electric fields and then maintained at zero power. This opens the possibility of creating a strain-tunable magnetic layer as a spin injector in a spintronic device with enhanced low power and permanent storage capabilities. Further device miniaturisation should provide opportunities for better understanding and control of the ferroelastic domains contribution to straintronics. Our findings should also motivate studies of sub-coercive behavior in (multi)ferroic systems, where we currently lack understanding in electric or elastic domains contributions to the observed hysteresis effects.

## Acknowledgements

Support of the Agence Nationale de la Recherche (hvSTR-ICTSPIN ANR-13-JS 04-0008-01, Labex NIE 11-LABX-0058-NIE, Investissement d'Avenir program ANR-10- IDEX-0002-02) and the technical support of the STnano clean room (H Majjad) are gratefully acknowledged.